\documentclass[12pt,superscriptaddress,nofootinbib]{revtex4-2} 

\usepackage{mathtools}

\usepackage{booktabs,makecell,array}
\newcolumntype{P}[1]{>{\raggedright\arraybackslash}p{#1}}

\usepackage{amsmath}  
\usepackage{amsfonts} 
\usepackage{graphicx} 
\usepackage{esvect}
\usepackage{mathtools}

\usepackage{xcolor}
\usepackage{appendix}
\usepackage{siunitx} 
\usepackage{mathrsfs}

\usepackage{placeins}

\usepackage[caption=false]{subfig} 

\usepackage{pbox}

\usepackage{mathtools}

\begin{document}

\title{Digitization Can Stall Swarm Transport: \\
Commensurability Locking in Quantized‑Sensing Chains}

\author{Caroline N. Cappetto}
\thanks{These authors contributed equally.}
\affiliation{Scripps College, Claremont, CA 91711, USA}

\author{Penelope Messinger}
\thanks{These authors contributed equally.}
\affiliation{Scripps College, Claremont, CA 91711, USA}

\author{Kaitlyn S. Yasumura}
\thanks{These authors contributed equally.}
\affiliation{Scripps College, Claremont, CA 91711, USA}

\author{Miro Rothman}
\thanks{These authors contributed equally.}
\affiliation{Pitzer College, Claremont, CA 91711, USA}

\author{Tuan K. Do}%
\affiliation{%
Department of Mathematics, University of California,\\ Santa Barbara, CA 93106, USA}%

\author{Gao Wang}
\affiliation{Wenzhou Institute, University of Chinese Academy of Sciences, Wenzhou, Zhejiang 325000, China}%

\author{Liyu Liu}
\affiliation{Human Phenome Institute, Fudan University, Shanghai 201203, China}%

\author{Robert H. Austin}
\affiliation{Department of Physics, Princeton University, Princeton, NJ 08544, USA}%

\author{Shengkai Li}
\thanks{Corresponding author: \texttt{shengkaili@princeton.edu}}
\affiliation{Department of Physics, Princeton University, Princeton, NJ 08544, USA}%

\author{Trung V. Phan}%
\thanks{Corresponding author: \texttt{tphan@natsci.claremont.edu}}
\affiliation{Department of Natural Sciences, Scripps and Pitzer Colleges, \\ Claremont Colleges Consortium, Claremont, CA 92110, USA}%

\begin{abstract}
We present a minimal model for autonomous robotic swarms in both one-dimensional and higher-dimensional spaces, where identical, field-driven agents interact pairwise to self-organize spacing and independently follow local gradients sensed through quantized digital sensors. We show that the collective response of a multi-agent train amplifies sensitivity to weak gradients beyond what is achievable by a single agent. We discover a fractional transport phenomenon in which, under a uniform gradient, collective motion freezes abruptly whenever the ratio of intra‑agent sensor separation to inter‑agent spacing satisfies a number‑theoretic commensurability condition. This commensurability locking persists even as the number of agents tends to infinity. We find that this condition is exactly solvable on the rationals --  a dense subset of real numbers -- providing analytic, testable predictions for when transport stalls. Our findings establish a surprising bridge between number theory and emergent transport in swarm robotics, informing design principles with implications for collective migration, analog computation, and even the exploration of number-theoretic structure via physical experimentation.
\end{abstract}

\date{\today}

\maketitle

\section{Introduction}

Inspired by decentralized swarm behavior in nature \cite{duan2023animal,tan2013swarm,cavagna2017dynamic}, robotic swarms offer scalable automation that continues working reliably even when some robots fail \cite{schranz2020swarm,winfield2006safety,bjerknes2013fault}. What makes robot swarms effective for engineering purposes is not only distributed decision-making \cite{vigelius2014multiscale,valentini2015efficient} but also their ability to self-organize movement \cite{zhou2022swarm,chung2018survey} -- sensing weak signals, aligning directions, and migrating as a coherent collective despite noise, failures, and limited communication -- hallmarks of natural swarms \cite{wadhams2004making,viscido2004individual,cavagna2014bird}. Because coordinated locomotion under minimal sensing -- often via quantized digital sensors \cite{iskakov2007influence,pu2018rotation,wang2021emergent,wang2022robots}, unlike natural sensing -- and simple spacing control \cite{brambilla2013swarm,endsley2017autonomous} is a core primitive for nearly every application, understanding when and how swarms move -- or stall -- in response to environment cues is a fundamental question of first importance in robophysics.

Here, we investigate the collective response of identical, field-driven agents -- arranged as one-dimensional chains \cite{apoorva2018motion,sinaasappel2025collecting} or as two-dimensional structures \cite{hernandez2024model} -- to a constant external drive, e.g. a uniform gradient. We do so through both \textit{in-silico} experiments and theoretical analysis. These active, autonomous agents interact pairwise via a damping constraint that enforces equal spacing, and each independently senses the local field using multiple digital sensors with quantized resolution to decide its future movement \cite{phan2021bootstrapped}. We see ``more is different'' \cite{anderson1972more}: once the agents are coupled, their collective response could outperform any individual, boosting sensitivity to weak gradients. Curiously, we find that the collective response can abruptly collapse to near-zero across parameter space, e.g. gradient strength and inter-agent spacing, revealing a fractal pattern which persists even for infinitely many agents. We explain this emergent transport phenomenon by mapping it to a number-theoretic commensurability problem \cite{ostlund1981incommensurate}, which we provide a complete solution on the rationals $\mathbb{Q}$ (a dense subset of the reals $\mathbb{R}$) consistent with our simulations. 

We introduce the mathematical model of the robot swarm in Section \ref{sec:model}, report simulation observations in one-dimensional spaces in Section \ref{sec:obs}, and present a theoretical explanation in Section \ref{sec:theo}. In Section \ref{sec:obs_higher}, we showcase some interesting findings in higher-dimensional spaces. Our results link number theory with robotic swarm behavior and identify concrete levers for reliable operation, while enabling physical probes of arithmetic structure. Commensurability locking phenomena have been observed and studied in condensed matter physics, with many potential for application \cite{petit2012commensurability,reichhardt2010commensurability,kasper2020simulating}. In principle, this could open a path to swarm-based analog computation, where collective flow stands in for electron currents -- but with greater controllability. 

\begin{figure*}[htbp]
\centering
\includegraphics[width=0.95\textwidth]{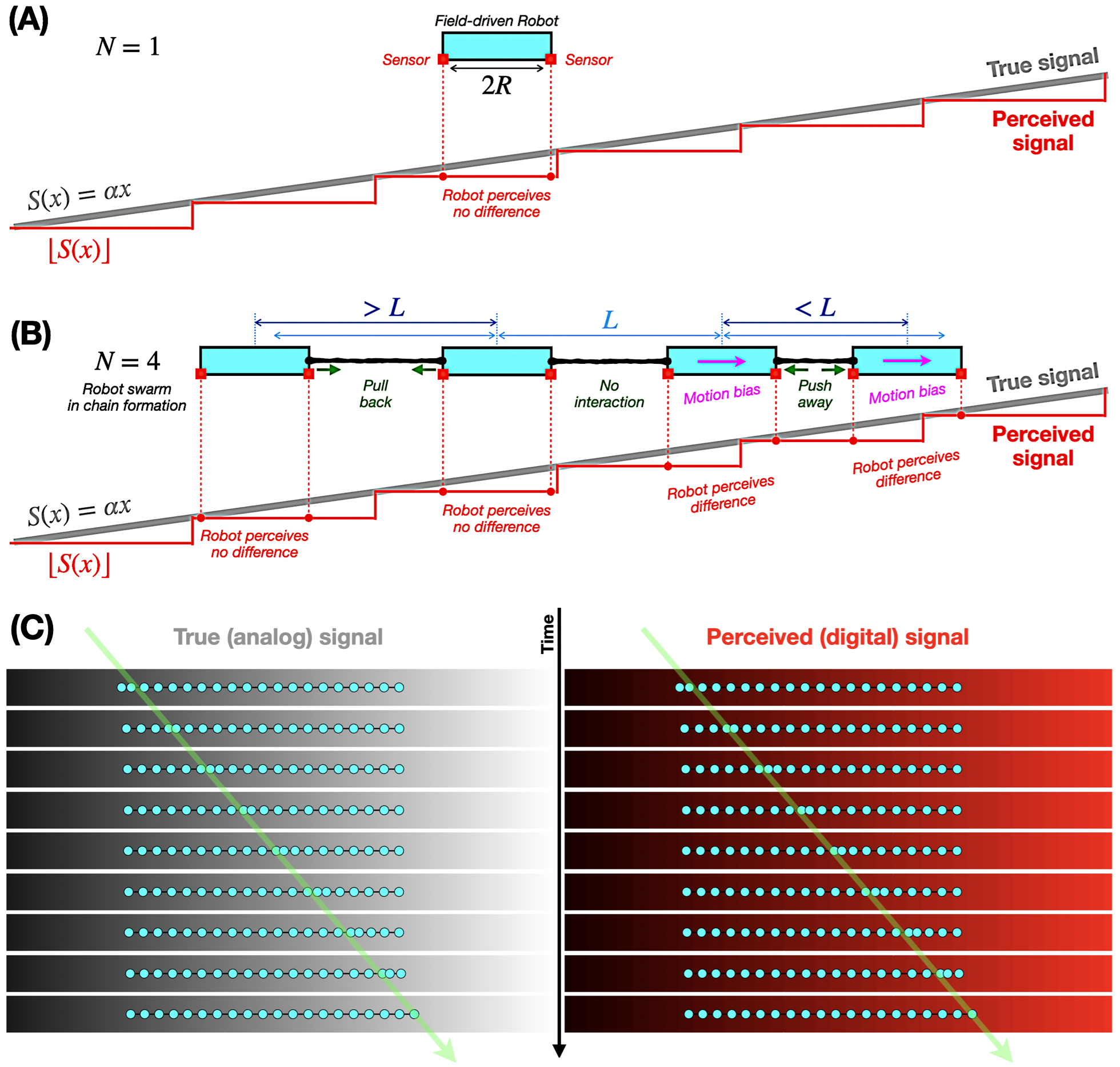}
\caption{\textbf{Field-driven robots on an environment with constant weak gradient.} Our illustrations are for one-dimensional systems; denote $N$ for the number of robots in the swarm. \textbf{(A)} The perceived signal received by digital sensors, due to quantization, might be identical and thus the agent ($N=1$) has no motion bias. \textbf{(B)} Inter-agent interactions hold the group loosely cohesive while permitting internal motion; a small subset of agents that perceive the signal differences can then have motion bias, thus drive the entire group ($N=4$) forward. \textbf{(C)} The \textit{crawling mode} of transport, in which a longitudinal traveling wave (green arrow) transmits information along the swarm. We show both the true signal and the perceived signal received be the quantized sensors. The snapshots are in chronological order, from top to bottom.}
\label{fig01}
\end{figure*}

\section{Robotic Swarm Model \label{sec:model}}

We start by formulating the mathematical model used for the robotic swarm in our study. For ease of exposition, we focus on the one-dimensional case, noting that the ideas extend naturally to higher dimensions. With $X_j(t)$ the center position of the $j$-th robot at time $t$ and $\Delta t$ the discrete time between consecutive time steps, we generate all robot trajectories via:
\begin{equation}
    X_j(t+\Delta t) - X_j(t) = \lambda_{\text{rand}} + \lambda_{\text{sens}} + \lambda_{\text{int}} \ .
    \label{eq:update_pos}
\end{equation}
Let us explain, separately, the three contributions (i.e. $\lambda_{\text{rand}}$, $\lambda_{\text{sens}}$, and $\lambda_{\text{int}}$)  to the total displacement:
\begin{itemize}
    \item Each robot moves with some stochasticity given by a Wiener increment:
    \begin{equation}
        \lambda_{\text{rand}}=(2 D \Delta t)^{1/2} \eta_j(t) \ ,
    \label{eq:rand}
    \end{equation}
    where $D$ is the effective diffusivity that quantifies its motion randomness. The factor $\eta_j(t)$ is an independent draw from a standard normal distribution, $\mathcal{N}(0,1)$, with independence across robots $j$ and time steps $t$.
    \item Every robot has a physical size $R$, with sensors mounted at its two ends, i.e. located at $X_j(t)\pm 1$. Each sensor returns a quantized reading of the local signal, so that if the analog input is $S(x)$, where $x$ is the position of the sensor, then the perceived signal is digitized into the floor value $\lfloor S(x) \rfloor$ (see Fig. \ref{fig01}A). Based on the perceived signals, the robot compares the two and biases its motion toward the higher value:
    \begin{equation}
        \lambda_{\text{sens}} = \kappa \frac{\left\lfloor S\left( X_j(t)+R \right)\right\rfloor - \left\lfloor S\left( X_j(t)-R \right)\right\rfloor}{2R} \Delta t \ ,
    \label{eq:sens}
    \end{equation}
    where $\kappa$ controls the magnitude of the directed drift induced by sensing (i.e. field-drive). After a standard spacetime rescaling, we nondimensionalize the model so that $R=1$ and $\kappa=1$ \cite{manteca2012mathematical,phan2024vanishing}.
    \item Robots interact pair-wise with fixed neighbors to maintain the swarm formation. The interaction pulls when separations exceed a distance $L$ and pushes when they fall below $L$ (see Fig. \ref{fig01}B), thus stabilizing the spacing at $L$ via a damping interaction:
    \begin{equation}
        \lambda_{\text{int}} = \mu \sum^{\text{neighbors of } j}_k \left( \left| \delta_{kj}(t)\right|-L \right) \frac{ \delta_{kj}(t)}{\left| \delta_{kj}(t)\right|} \Delta t \ ,
    \label{eq:int}
    \end{equation}
    where $\delta_{kj}(t)=X_k(t)-X_j(t)$ and $\mu$ is the damping coefficient that sets the rate at which pairwise spacing deviations (compared to $L$) are corrected. For a one-dimensional chain formation, the robots are labeled in order $j=1,2,3,...,N$, and interactions occur only between consecutive indices, i.e. the interaction graph is given by $\{ (j,j+1)\}^{N-1}_{j=1}$.
\end{itemize}
We can use Eq. \eqref{eq:rand}, Eq. \eqref{eq:sens} and Eq. \eqref{eq:int} to calculate position update with Eq. \eqref{eq:update_pos}, allowing us to simulate robotic swarm behavior. In this work, we only consider motion in a uniform gradient, i.e. $S(x)=\alpha x$. Together, the four constants
$$ \alpha \ , \  L \  , \ \mu \ , \ \text{and} \ D $$
parameterize the model. To keep neighboring robots separated beyond their physical size, we require $L \geq 2$. Moreover, by reflection symmetry $x \rightarrow -x$, $\alpha$ and $-\alpha$ are equivalent behavior (but in opposite directions), hence it suffices to consider $\alpha \geq 0$.

It can already be seen from Eq. \eqref{eq:sens} that when the environmental gradient is sufficiently weak -- the signal difference $2R\alpha$ between two sensors is below the quantized resolution $1$, thus \begin{equation}
    \alpha < 1/2R = 0.5
    \label{beyond_single_cap}
\end{equation}
-- there exist positions where the two sensors on the same robot perceive identical signals, and therefore no motion bias arises, i.e. $\lambda_{\text{sens}}=0$ (see Fig. \ref{fig01}A). In a robot swarm, some robots may occupy positions where their sensors perceive a difference; these agents are field-driven up-gradient, and their interactions -- intended to maintain the swarm’s formation -- can mediate this information and bias the collective motion (see Fig. \ref{fig01}B).

\begin{figure*}[htbp]
\centering
\includegraphics[width=\textwidth]{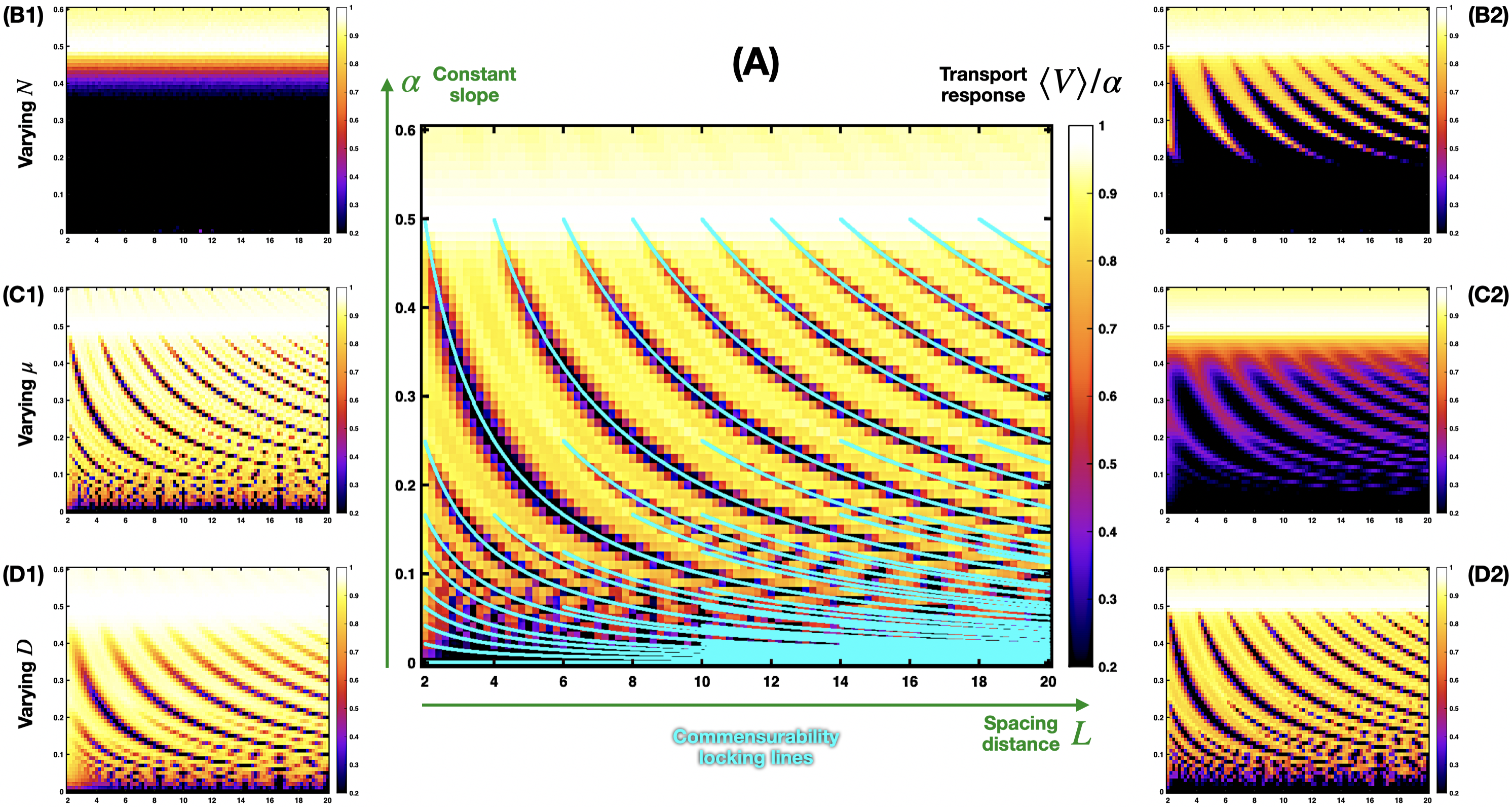}
\caption{\textbf{Simulation results for the transport response of different robot swarm chains.} For each specific choice for the number of robots $N$, the damping coefficient $\mu$, and the effective diffusivity $D$, we report the estimated transport response $\langle V \rangle /\alpha$ (the dark-to-light color gradient shows low-to-high values) for a range of constant slopes $\alpha \in [0,0.6]$ and spacing distances $L \in [2,20]$. \textbf{(A)} Baseline: $N=100$, $\mu=0.1$, $D=0.01$ (a long chain approximating the infinite-robot limit). Cyan lines mark loci in $(L,\alpha)$ where commensurability locking is predicted by theory (Sec.~\ref{sec:theo}). \textbf{(B1–B2)} Vary $N$ only: $N=1$ and $N=2$ (with $\mu=0.1$, $D=0.01$). \textbf{(C1–C2)} Vary $\mu$ only: $\mu=0.3$ and $\mu=0.01$ (with $N=100$, $D=0.01$). \textbf{(D1–D2)} Vary $D$ only: $D=0.1$ and $D=0.001$ (with $N=100$, $\mu=0.1$). All other parameters are held at the baseline when not being varied.}
\label{fig02}
\end{figure*}

\section{Fractional Stalling in One-Dimensional Space \label{sec:obs}}

Using \textit{MatLab 2025a} \cite{MATLAB-R2025a}, we implement the model in Section \ref{sec:model} to simulate one-dimensional robot swarms arranged in chain formations. We fix the number of robots $N$, damping $\mu$, and effective diffusivity $D$, and scan the parameter grid $(L,\alpha) \in [2,20]×[0,0.6]$ with discretizations $\Delta L=0.2$ and $\Delta \alpha = 0.1$. In all panels of Fig. \ref{fig02}, each pixel reports the values of the transport response (the dark-to-light color gradient shows low-to-high values) for a single simulation (with the time discretization $\Delta t=0.1$ and the total time duration $T=10^5$). The transport response is estimated from a time-averaging statistics as $$\langle V \rangle / \alpha \ \ \text{with} \ \  \langle V \rangle = \frac{\langle X(t=T) \rangle - \langle X(t=0) \rangle }T \ , \ $$
where $\langle X(t) \rangle$ is the center of mass location (space-average) of all robots at time $t$. The transport response equals $1$ in the ideal limit --robots driven solely by the bias in Eq. \eqref{eq:sens}, with analog (undigitized) sensing that reports the true signal. Interpreting the robots as responsive particles driven by a uniform field (the constant gradient), the emergent transport response serves as a conductivity-like coefficient (e.g. drift mobility).

We begin, in Fig. \ref{fig02}A, with a baseline swarm of $N=100$ robots (to approximate an infinite-long chain), with inter-robot damping coefficient $\mu=0.1$ and effective diffusivity $D=0.01$ for each robot. Compared to when there is only a single robot, i.e. $N=1$, as in Fig. \ref{fig02}B1, we find that collective effects enhance gradient sensitivity, yielding a substantially larger transport response for weak slopes $\alpha < 0.5$. The gradient sensitivity can be regained by increasing the number of robots in the swarm, e.g. adding one more robot ($N=2$) already creates new regions in $(L,\alpha)$ parameter space with significant transport responses (see Fig. \ref{fig02}B2). The larger the swarm, the closer the behavior is to Fig. \ref{fig02}A -- our infinite-chain baseline; in Appendix \ref{app:larger_swarm} we show more results, for the number of robots varying from $N=3$ to $N=8$. At these weak slopes, the swarm frequently exhibits a \textit{crawling mode}, in which traveling compression–expansion waves move along the up-gradient (see Fig. \ref{fig01}C and Supplementary Movie \textit{smovie01.mp4}).

It is important to note that, even for this baseline, there are still lines in the $(L,\alpha)$ parameter space along which the swarm loses sensitivity to the gradient. We attribute this collective transport stall to a commensurability locking effect, as conjectured in Section \ref{sec:theo}, whose predictions agree precisely with our simulations (see the cyan lines on Fig. \ref{fig02}A).

We also investigate the parameter influence by perturbing the baseline: we vary the damping coefficient $\mu$ and the diffusivity 
coefficient $D$ in relative to the baseline, as shown in Fig. \ref{fig02}C and Fig. \ref{fig02}D, respectively. Higher $\mu$ or $D$ amplifies the sensitivity (see Fig. \ref{fig02}C1 and Fig. \ref{fig02}D1), whereas lower $\mu$ or $D$ diminishes it (see Fig. \ref{fig02}C2 and Fig. \ref{fig02}D2).

\section{A Number-Theoretic Explanation \label{sec:theo}}

The following is our conjecture to explain the stalling phenomena reported in Section \ref{sec:obs}. The collective transport of the robot swarm can stall because, for certain values of parameters $\alpha$ and $L$, there exists a spatial distribution of agents, equally spaced by $L$, in which none of the agents detects any difference between the digitized values perceived by its sensors -- even in the limit of infinitely-many agents $N\rightarrow \infty$. Therefore, the set $\mathcal{S}$ of all $(\alpha,L)$ for which this situation may occurs can be described as a number-theoretic commensurability problem:
\begin{equation}
\begin{split}
& \mathcal{S} = \Big\{  (\alpha,L) : \alpha \in [0,0.5) , L \in [2,+\infty) \Big| 
\\
& \ \ \ \ \ \ \ \ \exists x\in \mathbb{R}, \forall (m,n) \in \mathbb{Z}^2 : x+ \frac{m}{\alpha} + n L \notin [-1,+1] \Big\}  
\end{split}
\label{number_problem} 
\end{equation}
-- see Appendix \ref{app:commensurability} for a full derivation. While we are still not able to completely solve this problem, we can if we focus on its rational subset:
\begin{equation}
    \mathcal{S}_{\mathbb{Q}}=\mathcal{S}\cap  \mathbb{Q}^2 \ ,
\end{equation}
which is dense in $\mathcal{S}$ -- meaning any $(\alpha,L)$ in $\mathcal{S}$ can be approximated arbitrarily closely by rational pairs in $\mathcal{S}_{\mathbb{Q}}$. Intuitively, restricting to rational parameters loses no generality for approximation, i.e. results proved on $\mathcal{S}_{\mathbb{Q}}$ should extend to all of $\mathcal{S}$ by continuity.

Let us write $\alpha=a_1/b_1$ and $L=a_2/b_2$ as irreducible fractions, which means $(a_i,b_i)\in \mathbb{Z}^2$ and gcd$(a_i,b_i)=1$ with $i=1,2$. As a corollary of the Euclid algorithm \cite{lehmer1938euclid}, we have the following set-equality:
\begin{equation}
    \{ma+nb : (m,n)\in \mathbb{Z}^2\} = \{k \times\text{ gcd}(a,b) : k\in \mathbb{Z}\}.
\end{equation}
By using this equality on the numerator of
\begin{equation}
\frac{m}{\alpha}+nL=\frac{mb_1}{a_1}+\frac{na_2}{b_2} = \frac{mb_1b_2+na_1a_2}{a_1b_2} \  ,  
\end{equation}
we see that:
\begin{equation}
\begin{split}
\left\{\frac{m}{\alpha}+nL : (m,n)\in \mathbb{Z}^2\right\} = 
\left\{ k \times \frac{\text{ gcd}(b_1b_2,a_1a_2)}{a_1b_2} : k\in \mathbb{Z} \right\} \ .
\end{split}
\end{equation}
Denote gcd$(b_1b_2,a_1a_2)/(a_1b_2)=c$, then:
\begin{itemize}
    \item If $c>2 $ then any value in the interval $(1,c-1)$ satisfies the condition in $|...\}$ of Eq. \eqref{number_problem}, thus $(\alpha,L)=(a_1/b_1,a_2/b_2) \in \mathcal{S}_{\mathbb{Q}}$.
    \item If $c \le 2$ then because the length of the interval $[-1,1]$ equals to $2$, such $(\alpha,L)$ will not land on $\mathcal{S}$.
\end{itemize}
Fig. \ref{fig02}A shows a reasonable agreement between the values of $(\alpha,L)$ that exhibit collective stalling and points in our solution set:
\begin{equation}
    S_{\mathbb{Q}} = \left\{(\alpha,L)= \left( \frac{a_1}{b_1},\frac{a_2}{b_2} \right) : \frac{\text{ gcd}(b_1b_2,a_1a_2)}{a_1b_2} >2  \right\} \ ,
\label{eq:locking}
\end{equation}
which can be generated numerically by scanning through integer values of $a_1$, $a_2$, $b_1$, and $b_2$.

There are many features of Fig. \ref{fig02} that our simple number-theoretic theory has yet to explain, such as the distribution of peaks and dips in the transport response. We also observe rich behavior at large slopes ($\alpha \geq 0.5$) -- we report our observations in Section \ref{app:large_slopes}.

\begin{figure*}[htbp]
\centering
\includegraphics[width=\textwidth]{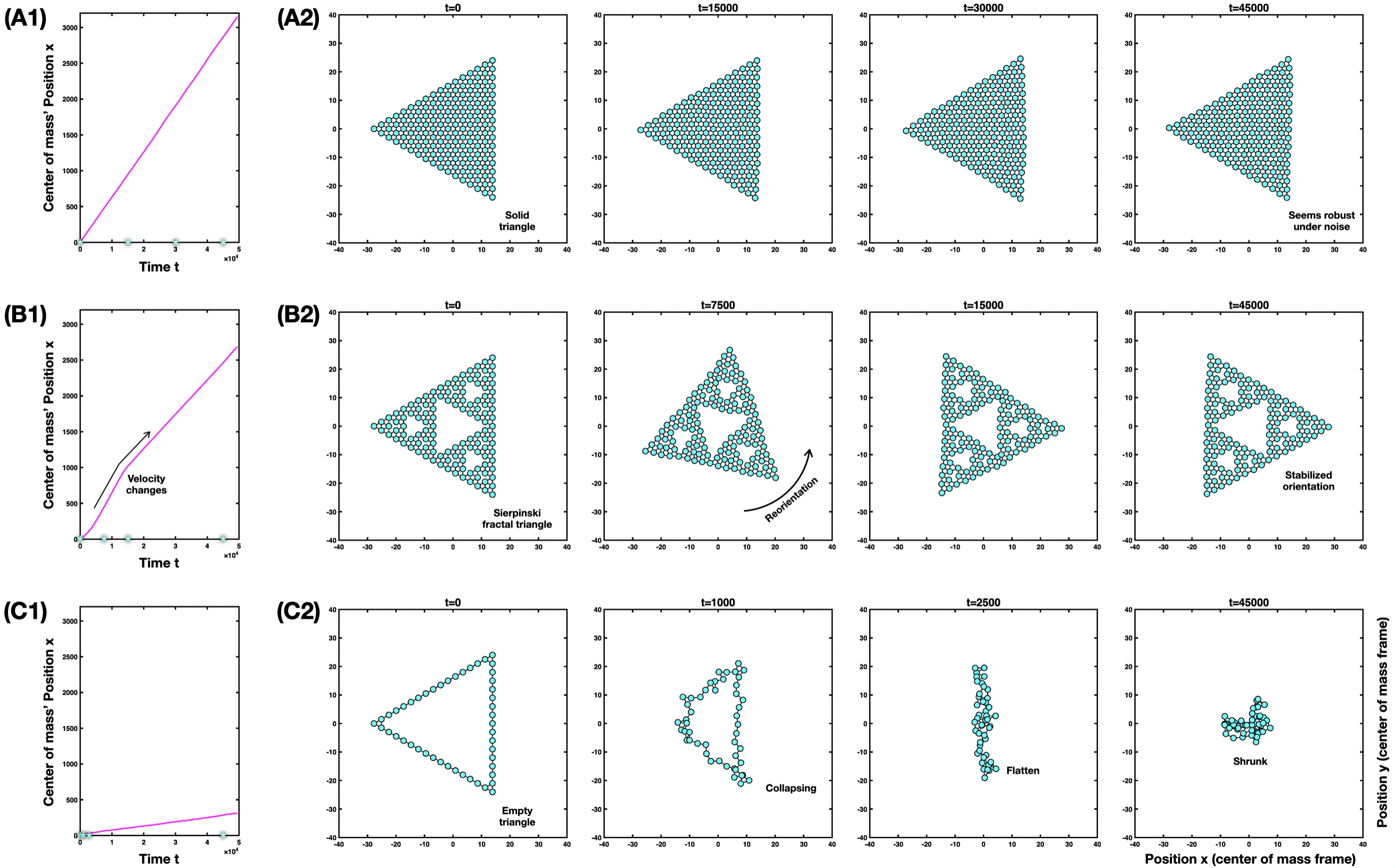}
\caption{\textbf{Simulation results for different formations of robot swarms in two-dimensional space.} We use the same parameters, only change the number of robots and formation topologies: a solid triangle in \textbf{(A)}, a Sierpinski fractal triangle in \textbf{(B)}, and an empty triangle in \textbf{(C)}. We report the time evolution of the center of mass $x$-position in \textbf{(A–C1)} and the corresponding swarm configurations in \textbf{(A–C2)}. Cyan stars in \textbf{(A–C1)} indicate the times of the snapshots in \textbf{(A–C2)}. All panels in \textbf{(A–C2)} use the same axis labels, so we show them only on the bottom-right panel.}
\label{fig03}
\end{figure*}

\section{Some Observations in Higher-Dimensional Spaces \label{sec:obs_higher}}

This section explores phenomena that emerge in higher-dimensional generalizations of the robotic swarm, using the same model as in Section \ref{sec:model} for simulations. The gradient always points along the $x$-direction, and the signal is otherwise uniform transversely -- the $y$-direction in two-dimensional space and the $xy$-plane in three-dimensional space. For simplicity, we assume the robots are always oriented with their sensors aligned along the 
$x$-direction. We use the same parameters across all simulations and dimensions, i.e. $\alpha=0.1$, $L=3$, $\mu=0.1$, and $D=0.01$; the only change is the swarm formation (number of robots and topology).

In two-dimensional space, we consider different triangle formations: a solid triangle (see Fig. \ref{fig03}A), a Sierpinski fractal triangle (see Fig. \ref{fig03}B), and an empty triangle (see Fig. \ref{fig03}C). In the robophysics literature, the first and last structures are known as \textit{active solids} \cite{hernandez2024model} and \textit{active polymers} \cite{sinaasappel2025collecting}; here we also consider \textit{active fractals} -- the middle structures. Active solids are very rigid; active fractals are more deformable; active polymers are the easiest to deform. We observed from our simulations that the collective transport response of the solid triangle swarm is the highest and that of the empty triangle swarm is the lowest (see Fig. \ref{fig03}A-C1). In the center of mass frame, over a total time $T=5\times 10^4$
(with time step $\Delta t=0.1$), the solid triangle swarm exhibits little rotation (see Fig. \ref{fig03}A2). By contrast, the Sierpinski fractal triangle swarm reorients and its transport response decreases (see Fig. \ref{fig03}B2). Our interpretation is that the reorientation establishes persistent internal constraints, diverting power into maintaining internal order rather than coherent up-gradient motion, so the emergent collective transport slows down. The hollow triangle collapses, flattening in the transverse direction then shrinking (see Fig. \ref{fig03}C2). All these behaviors are shown in Supplementary Movies \textit{smovie02.mp4}, \textit{smovie03.mp4}, and \textit{smovie04.mp4}. Similar collapsing behavior can also be seen in three dimensions for an \textit{active ``fuzzy sphere''} \cite{madore1992fuzzy,vo2024size} (arranged in a buckyball C$_{60}$ formation): the structure first ``pancakifies'' onto the transverse plane and then contracts (see Supplementary Movies \textit{smovie05.mp4}). The collapsed buckyball also exhibits a very clear \textit{crawling motion} -- appearing to expand and contract in the center of mass frame along the $x$-direction -- as it climbs up the gradient (see Appendix \ref{app:3Dsim}).

\section{Discussion}

We have found that sensing quantization can make an infinite chain of field-driven robots stall in a one-dimensional constant drive, when the distance between sensors and robot spacing with this swarm satisfy the commensurability locking condition Eq. \eqref{eq:locking}. In two- and three-dimensional spaces, depending on the swarm topology, the structure can reorient, collapse, and alter its collective transport properties. These physical properties are fundamental to the transport of robotic swarms using digital sensing, in the limit where individual capabilities become ineffective and collective effects dominate.

\begin{figure*}[htbp]
\centering
\includegraphics[width=\textwidth]{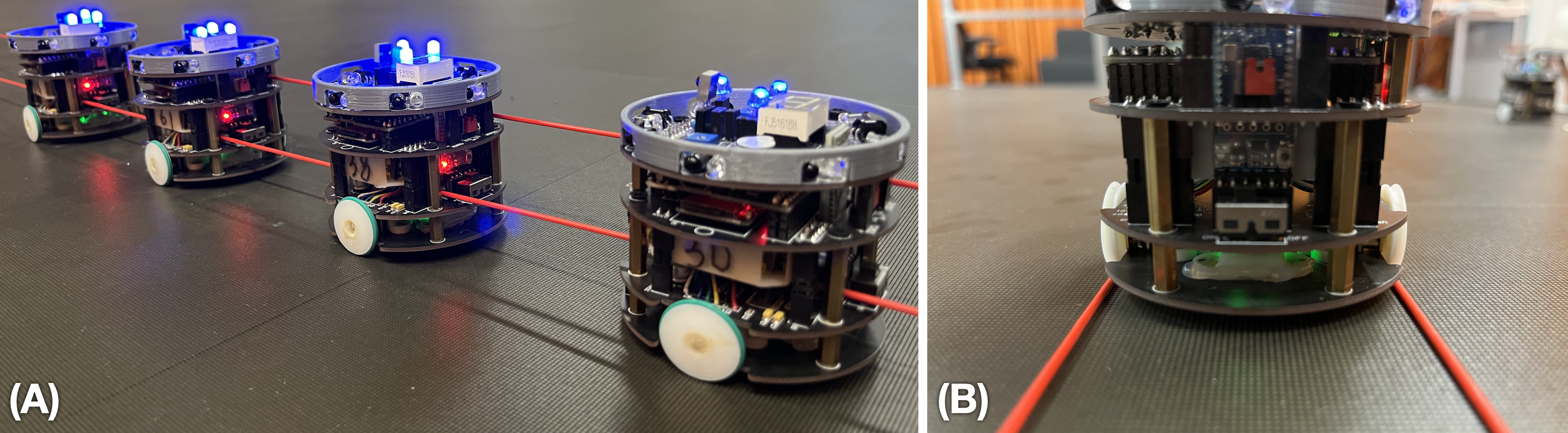}
\caption{\textbf{Experimental setups for a one-dimensional robotic swarm in chain formation.} We are evaluating multiple methods for executing these experiments. \textbf{(A)} A chain of robots constrained by a guiding line passing through each unit to keep them collinear. \textbf{(B)} A robot placed on two parallel rail tracks. We show a preliminary experiment for robots running on tracks in the Supplementary Movie \textit{smovie06.mp4}. Infrared sensors on each robot determine pairwise distances, which set the interaction strength.}
\label{fig04}
\end{figure*}

It is important to note that here we have only scratched the surface of a potentially deep topic. There are many interesting features of the data that remain unexplained, such as the local peaks and dips in the transport response beside commensurability locking (as shown in Fig. \ref{fig02} and Fig. \ref{figS01}). There are many further directions to explore, such as considering different swarm formation structures, performance robustness to failed or deactivated robots, collective response to time-varying drives or adaptive spacing to deliberately break commensurability, and replacing fixed connections to active ones in which direct interacting relations evolve dynamically via Thiessen/Dirichlet–Voronoi tessellations \cite{burrough2015principles} (which preliminary findings have been reported in \cite{van2021swarm}). The randomness of robot motion can be made more biologically relevant by using an adaptive random walk \cite{nguyen2024remark} instead of a simple random walk as described in Eq. \eqref{eq:rand}; this feature could have future implications for bioengineering. Physical experiments are underway and will be reported in future work (see Fig. \ref{fig04}).

\section{Acknowledgements}

This work was supported by the National Natural Science Foundation of China (Nos.T2350007, 12404239, 12174041) and the US National Science Foundation (PHY-1659940 and PHY-1734030). S.L. was supported by the National Science Foundation, through the Center for the Physics of Biological Function (PHY-1734030) in Princeton University. We acknowledge useful discussions with Truong H. Cai, Huy D. Tran, Khang V. Ngo, and Van H. Do.

\appendix 

\section{Transport Response of Finite Swarms \label{app:larger_swarm}}

\begin{figure*}[htbp]
\centering
\includegraphics[width=\textwidth]{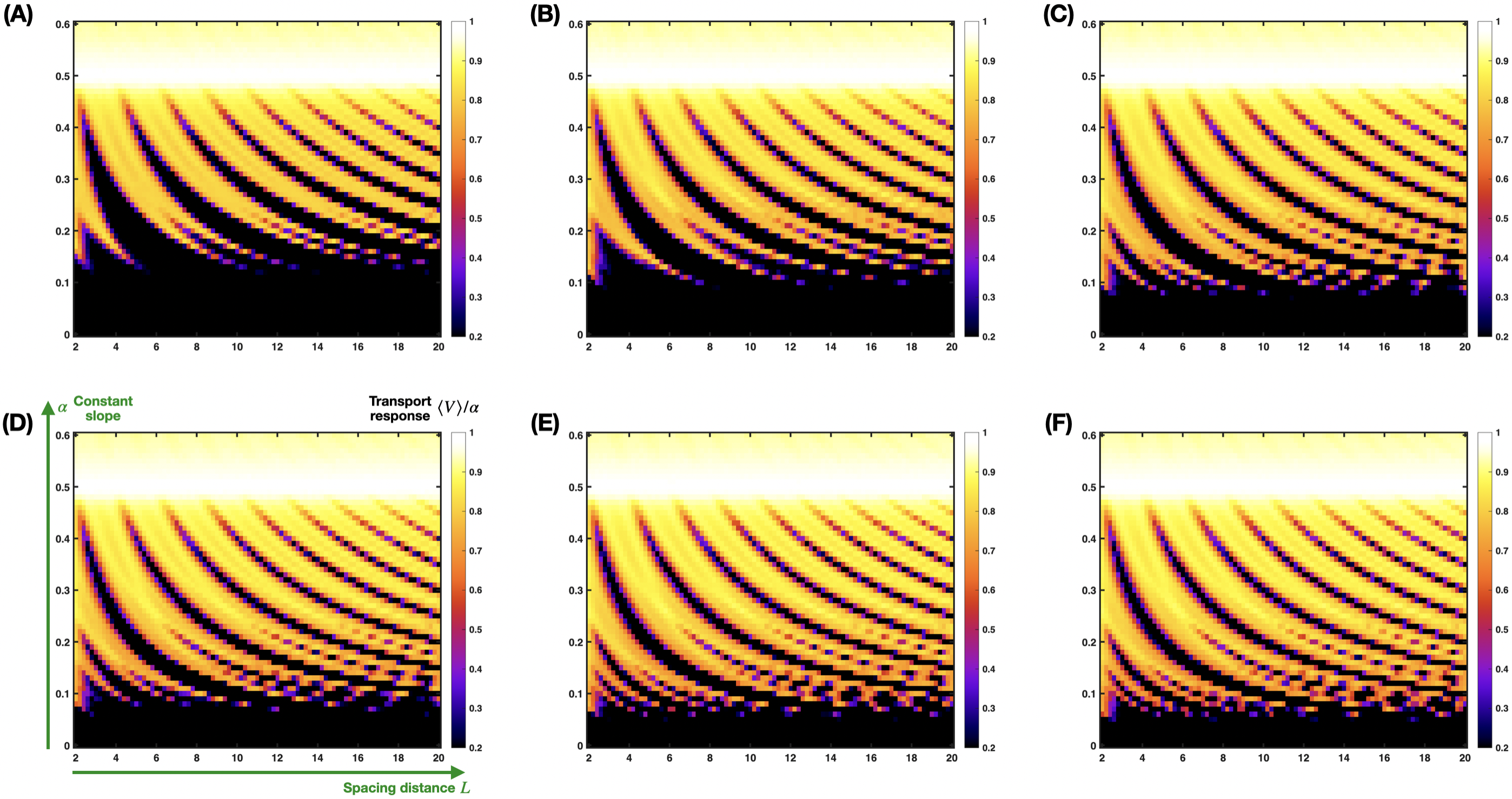}
\caption{\textbf{More simulation results for the transport response of different robot swarm chains.} Here, every simulation's details are the same with the baseline in Fig. \ref{fig02}, except that we vary the number of robots in the chain. \textbf{(A)} $N=3$. \textbf{(B)} $N=4$. \textbf{(C)} $N=5$. \textbf{(D)} $N=6$. \textbf{(E)} $N=7$. \textbf{(F)} $N=8$. All
panels use the same axis labels, so we show them only on the bottom-left panel.}
\label{figS02}
\end{figure*}

We perform simulations similar to those in Sec \ref{sec:obs}, with the number of robots $N$ goes from $3$ to $8$. We report our finding in Fig. \ref{figS02}.

\section{The Derivation for the Number-Theoretic Commensurability Problem \label{app:commensurability}}

The environment field is linear, i.e. $S(x)=\alpha x$. The robot compares two sensors reading at symmetric offsets, i.e. $\pm 1$. The perceived quantization of each sensor maps true analog values to integers, i.e. via a floor function. In weak field limit, i.e. $\alpha < 0.5$, two sensors on the same robot only disagree when the line-segment between them cross through a digitization boundary. Given that the locations of these boundaries are at integers of $S(x)$, i.e. $x=m/\alpha$ with $m \in \mathbb{Z}$, a robot -- whose center position is at $X_j$ -- can perceive a difference between its sensors only when 
\begin{equation}
    X_j \in \bigcup_{m \in \mathbb{Z}} \left[ \frac{m}{\alpha}-1,\frac{m}{\alpha}+1 \right] \ .
\end{equation}
Equivalently, the robot perceives no difference iff
\begin{equation}
    X_j + \frac{m}{\alpha} \notin [-1,1] \ \forall m \in \mathbb{Z} \ .
\end{equation}
To extend this condition to the collective of all robots in an infinite chain, we change $X_j \rightarrow X_j + n L$ (where $L$ is the formation spacing between nearby robots) and let the value $n$ scan through all integers, i.e.
\begin{equation}
    X_j + \frac{m}{\alpha} + nL \notin [-1,1] \ \forall (m,n) \in \mathbb{Z}^2 \ .
\end{equation}
This is essentially the condition presented in Eq. \eqref{number_problem}. The parameter restrictions are $\alpha \in [0,0.5)$ since we are investigating weak gradient, and $L \in [2,\infty)$ since we prevent the physical overlaps between nearby robots.

\section{Features of Transport Response for Large Slope $\alpha \geq 0.5$ \label{app:large_slopes}}

\begin{figure*}[htbp]
\centering
\includegraphics[width=\textwidth]{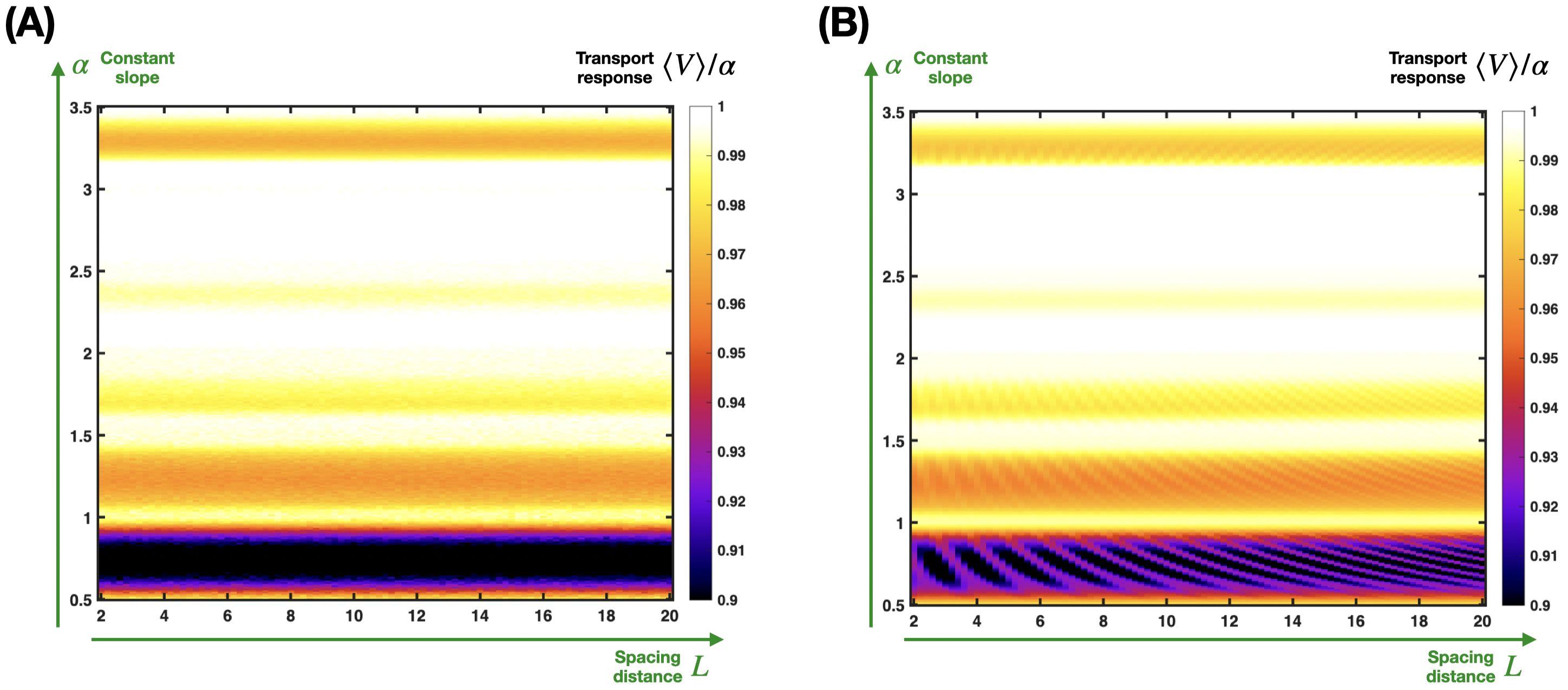}
\caption{\textbf{Simulation results for the transport response at large slope values $\alpha \geq 0.5$.} These experiments are done with the same damping coefficient $\mu=0.1$ and the effective diffusivity $D=0.01$, for the slope range $\alpha=[0.5,3.5]$ and the spacing distance range $L\in[2,20]$. \textbf{(A)} A single robot $N=1$. \textbf{(B)} A long chain of many robots $N=100$, approximating the infinite-$N$ limit.}
\label{figS01}
\end{figure*}

We perform simulations similar to those in Sec \ref{sec:obs}, with the constant slope $\alpha$ range goes from $0.5$ to $3.5$. For a single robot ($N=1$), there are irregular dips in the transport response (see Fig. \ref{figS01}A). For $N=100$ robot swarms, these dips do not disappear; instead, new peak features emerge (see Fig. \ref{figS01}B).

\section{Active ``Fuzzy Sphere'' Simulation \label{app:3Dsim}}

\begin{figure*}[htbp]
\centering
\includegraphics[width=\textwidth]{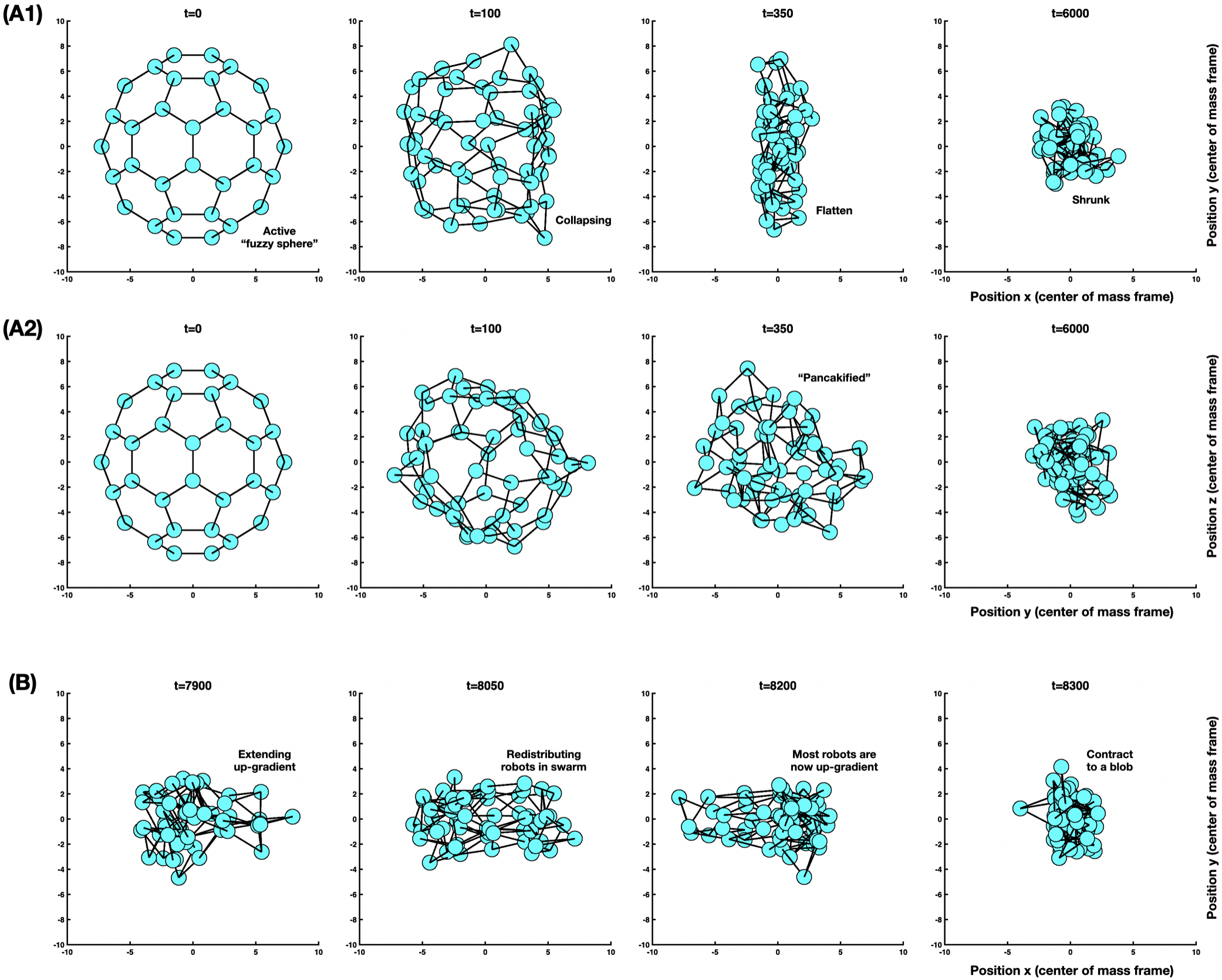}
\caption{\textbf{Simulation results for the buckyball formation of a robot swarm in three-dimensional space.} We use the same parameters with simulations done in Fig. \ref{fig03}, only changing the number of robots and formation topologies. \textbf{(A)} The time evolution of the robot swarm in the center of mass frame, showing the collapsing process, for \textbf{(A1)} the $xy$ plane and \textbf{(A2)} the (transverse) $yz$ plane. \textbf{(C)} The \textit{crawling motion} of the collapsed blob, in which clear alternating expansion and contraction can be seen, is shown in the center of mass frame for the $xy$ plane. All panels in \textbf{(A)}, \textbf{(B)}, and \textbf{(C)} use the same axis labels, so we show them only on the right-most panel.}
\label{fig05}
\end{figure*}

We show the behavior of the robot swarm with buckyball formation in three-dimensional space, in the swarm's center of mass frame, in Fig. \ref{fig05}. 

\bibliography{main}
\bibliographystyle{apsrev4-2}

\end{document}